\documentclass[reprint, superscriptaddress, amsmath,amssymb, aps, ]{revtex4-2}

\usepackage{acro}
\DeclareAcronym{qml}{
	short=QML,
	long=Quantum Machine Learning,}
\DeclareAcronym{tn}{
	short=TN,
	long=Tensor Network,
	long-plural=s,}
\DeclareAcronym{annni}{
	short=ANNNI,
	long=Axial Next-Nearest-Neighbor Ising,}
\DeclareAcronym{qcnn}{
	short=QCNN,
	long=Quantum Convolutional Neural Network,
	long-plural=s,}
\DeclareAcronym{qc}{
	short=QC,
	long=Quantum Circuit,
	long-plural=s,}
\DeclareAcronym{qae}{
	short=QAE,
	long=Quantum Autoencoder,
	long-plural=s,}
\DeclareAcronym{pqc}{
	short=PQC,
	long=Parametrized Quantum Circuit,
	long-plural=s,}
\DeclareAcronym{dmrg}{
	short=DMRG,
	long=density matrix renormalization group,
}
\DeclareAcronym{mps}{
	short=MPS,
	long=Matrix Product State,
	long-plural=s,}
\DeclareAcronym{fss}{
	short=FSS,
	long=finite-size scaling,
}
\DeclareAcronym{ml}{
	short=ML,
	long=Machine Learning,
}
\DeclareAcronym{vqe}{
	short=VQE,
	long=Variational Quantum Eigensolver,
}
\DeclareAcronym{kt}{
	short=KT,
	long=Kosterlitz-Thouless,
}
\DeclareAcronym{bcft}{
	short=BCFT,
	long=Boundary Conformal Field Theory,
}
\DeclareAcronym{cnn}{
	short=CNN,
	long=Convolutional Neural Network,
	long-plural=s,
}
\DeclareAcronym{rmse}{
	short=RMSE,
	long=Root Mean Squared Error,
}
\DeclareAcronym{cft}{
	short=CFT,
	long=Conformal Field Theory,
}
\DeclareAcronym{pt}{
	short=PT,
	long=Pokrovsky-Talapov,
}
\DeclareAcronym{mpd}{
	short=MPD,
	long=Matrix Product Disentangler,
}
\DeclareAcronym{hc}{
	short=HC,
	long=Hybrid Classifier,
	long-plural=s,
}
\DeclareAcronym{iqp}{
	short=IQP,
	long=Instantaneous Quantum Polynomial,
}

\usepackage{graphicx}
\usepackage{subcaption}
\usepackage{ragged2e}

\usepackage{mathtools}
\usepackage{physics}

\usepackage{hyperref}

\usepackage[dvipsnames]{xcolor}

\usepackage{tikz}
\usetikzlibrary{quantikz2}
\usetikzlibrary{fit,backgrounds,positioning}

\usepackage{xspace}

\begin{document}

\definecolor{qtiblue}{RGB}{0, 97, 174}
\definecolor{qtiteal}{RGB}{0,124,157}
\definecolor{qtigreen}{RGB}{0, 150, 140}

\def\linefa{Pokrovsky-Talapov\xspace} \def\linefp{Kosterlitz-Thouless\xspace}

\preprint{APS/Z2-preparation}

\title{Exploring the phase diagram of the quantum one-dimensional ANNNI model}

\author{M. Cea}
\affiliation{Max-Plank-Institut für Quantenoptik, Hans-Kopfermann-Str. 1, D-85748 Garching, Germany}
\affiliation{Munich Center for Quantum Science and Technology (MCQST), Schellingstr. 4, D-80799 München}

\author{M. Grossi}
\affiliation{European Organization for Nuclear Research (CERN), Geneva 1211, Switzerland}

\author{S. Monaco}
\thanks{Corresponding author.}
\email{saverio.monaco@desy.de}
\affiliation{RWTH Aachen University, 52062 Aachen, Germany}
\affiliation{Deutsches Elektronen-Synchrotron (DESY), D-22607 Hamburg, Germany}

\author{E. Rico}
\affiliation{European Organization for Nuclear Research (CERN), Geneva 1211, Switzerland}
\affiliation{Department of Physical Chemistry and EHU Quantum Center, University of the Basque Country UPV/EHU, Box 644, 48080 Bilbao, Spain}
\affiliation{Donostia International Physics Center, 20018 Donostia-San Sebastián, Spain}
\affiliation{IKERBASQUE, Basque Foundation for Science, Plaza Euskadi 5, 48009 Bilbao, Spain}

\author{L. Tagliacozzo}
\affiliation{Institute of Fundamental Physics IFF-CSIC, Calle Serrano 113b, Madrid 28006, Spain}

\author{S. Vallecorsa}
\affiliation{European Organization for Nuclear Research (CERN), Geneva 1211, Switzerland}

\date{\today}

\begin{abstract}
	This work presents a circuit-level integration of Quantum Machine Learning
	(QML) and Tensor Networks (TNs) in the context of the one-dimensional Axial
	Next-Nearest-Neighbor Ising (ANNNI) model with a transverse field. The study
	concretely connects these two paradigms: ground states of the model are
	obtained via the density matrix renormalization group (DMRG) as Matrix
	Product States (MPSs) and converted into quantum circuits, which are then
	used as input to a hybrid quantum-classical classifier and a Quantum
	Autoencoder (QAE) by prepending the state-preparation circuit to the
	trainable ansatz. This construction directly addresses a key bottleneck of
	purely quantum workflows, namely the need for expensive, gradient-based
	optimization on hardware for every input state, by delegating state
	preparation to classical TN methods, which are scalable by construction. The
	result is a scalable and hardware-compatible pipeline for the task of
	studying spin systems.
\end{abstract}

\maketitle

\section{\label{sec:intro}Introduction}

\ac{qml} and \acp{tn} are both prominent tools of quantum computation that share
common ground in their applications to complex quantum systems, yet they are
most often studied and benchmarked against each other rather than used in a
synchronized, complementary fashion. \ac{qml} leverages parametrized quantum
circuits to process and learn from data, with the potential for computational
advantages over classical methods, though it faces practical challenges in the
current non-fault-tolerant era, including noise, limited qubit counts, and
trainability issues such as barren plateaus. \acp{tn}, on the other hand,
provide a powerful classical framework for efficiently representing and
manipulating high-dimensional quantum states by exploiting their entanglement
structure, finding widespread use in condensed matter physics, quantum
information theory, and \ac{ml}.

In this paper, we exploit both paradigms across the various stages of the
overall algorithm's construction. While some literature explores comparing these
two approaches, only a few examples focus on combining them. Indeed, in the
context of variational quantum algorithms, the authors of
Refs.~\cite{29_ManuelRef_mpsvsPqc,Barratt_2021} explored the use of \acp{pqc}
for calculating diverse properties of complex many-body systems, describing the
benefits of this method relative to the use of \acp{tn}. In
Ref.~\cite{Huggins_2019}, the authors describe a joint optimization framework
utilizing both \acp{tn} and \acp{pqc}; the benefits of this style of joint
training were predicted to give improved performance in \ac{qml}, and this was
eventually verified numerically in Ref.~\cite{Dborin_2022}.

Finally, the authors of Ref.~\cite{Synergistic} proposed a synergistic
pretraining of quantum circuits via \acp{tn} applicable to a diverse range of
circuit architectures and learning tasks, which was predicted to yield benefits
in performance and trainability within general \ac{ml} tasks.

More broadly, avoiding the optimization loop on quantum hardware altogether has
recently emerged as a promising direction for scaling \ac{qml} beyond what is
reachable with today's noisy, resource-limited devices, since running the many
circuit evaluations required by gradient-based training directly on hardware
does not scale with system
size~\cite{recio2025trainclassical,li2025lwpp,monaco2025symbolicpp}. \acp{tn}
offer a complementary, and often more scalable, route to the same goal for
state-preparation tasks: rather than approximating the cost landscape of a fixed
ansatz, they provide efficient classical representations of the target states
themselves, from which a hardware-ready circuit can be constructed directly,
without any optimization loop at all.

In our work, we realize this combination at the circuit level. Ground states of
the \ac{annni} model are first computed classically as \acp{mps} via \ac{dmrg},
and are then converted into quantum circuit ans\"{a}tze using the \ac{mpd}
encoding from Ref.~\cite{ran2020}. These state-preparation circuits are directly
prepended to a quantum-classical \ac{hc} and the \ac{qae} architecture, forming
fully gate-based pipelines that can in principle be executed on real quantum
hardware.
Specifically, we focus on the phase diagram reconstruction for the \ac{annni}
model.

The \ac{annni} model (Fig.~\ref{fig:annni})
\cite{elliott1961phenomenological,fisher1980infinitely,selke1988annni,chakrabarti2008quantum}
with transverse field consists of ferromagnetic interactions, $J_1>0$, between
nearest neighbors and antiferromagnetic interactions, $J_2<0$, between
next-nearest neighbors, competing with each other. The Hamiltonian is written as
\begin{align}
	H_\text{ANNNI}= & -J_1\sum_{i=1}^{N-1}\sigma^{x}_{i}\sigma^{x}_{i+1}-J_{2}\sum_{i=1}^{N-2}\sigma^{x}_{i}\sigma^{x}_{i+2}\nonumber \\
	                & -B\sum_{i=1}^{N}\sigma^{z}_{i},
	\label{eq:H}
\end{align}
which we can rewrite in terms of the dimensionless ratios $\kappa=-J_2/J_1$
(called the \textit{frustration parameter}) and $h=B/J_1$.

\begin{figure}
	\centering
	\newcommand{\spindown}{
	\begin{tikzpicture}[baseline=(c.center)]
		\node[draw=black, thick, circle, minimum size=.5cm, inner sep=0pt, fill=qtiblue!50] (c) at (0,0) {};

		\draw[
		-{Triangle[width=8pt,length=8pt]},
		line width=4pt,
		color=qtiblue!40!gray
		] (0, +.25) -- ++(0,-.5); 

		\node[draw=black, thick, circle, minimum size=.5cm, inner sep=0pt] (c) at (0,0) {};

	\end{tikzpicture}
}
\newcommand{\spinup}{
	\begin{tikzpicture}[baseline=(c.center)]
		\node[draw=black, thick, circle, minimum size=.5cm, inner sep=0pt, fill=qtigreen!50] (c) at (0,0) {};

		\draw[
		-{Triangle[width=8pt,length=8pt]},
		line width=4pt,
		color=qtigreen!40!gray
		] (0, -.25) -- ++(0,+.5); 

		\node[draw=black, thick, circle, minimum size=.5cm, inner sep=0pt] (c) at (0,0) {};

	\end{tikzpicture}
}
\begin{tikzpicture}[
	connection/.style={
	->,
	line width=2pt,
	black!40!white,
	opacity=1,
	>={Triangle[length=4, width=4]}
	},
	magfield/.style={
	->,
	line width=2pt,
	black!40!white,
	opacity=1,
	>={Triangle[length=4, width=4]}
	}
	]
	\pgfdeclarelayer{background}
	\pgfdeclarelayer{main}
	\pgfdeclarelayer{foreground}
	\pgfsetlayers{background,main,foreground}

	\begin{pgfonlayer}{main}
		\node (spin11) at (0, 0) {\spinup};
		\node (spin21) at ([xshift=.5cm]spin11.east) {\spindown};
		\node (spin31) at ([xshift=.5cm]spin21.east) {\spindown};
		\node (spin41) at ([xshift=1cm]spin31.east) {\spinup};

		\draw[magfield] (-1, .5) -- coordinate (mag) ++(0, 0.75);
		\node at ([xshift=.3cm]mag) {$\mathbf{B}$};

		\draw[connection]
		([yshift=+.2cm]spin11) to[out=70, in=110] (spin21);
		\draw[connection]
		([yshift=+.2cm]spin11) to[out=-70, in=250] (spin31);

		\node at ([xshift=.05cm, yshift=-.9cm]spin31) {$\boldsymbol{J_2}$};
		\node at ([xshift=.25cm, yshift=.9cm]spin11) {$\mathbf{J_1}$};

	\end{pgfonlayer}

	\begin{pgfonlayer}{background}
		\draw[gray, line width=0.1pt] (spin11) -- (spin31);
		\draw[dashed, gray, line width=0.1pt] (spin41) -- (spin31);
	\end{pgfonlayer}
\end{tikzpicture}
	\caption{\justifying Sketch of the one-dimensional \ac{annni} model. Spins
		interact via a ferromagnetic nearest-neighbor coupling $J_1>0$, an
		antiferromagnetic next-nearest-neighbor coupling $J_2<0$, and are
		subject to a transverse magnetic field $B$, as in Eq.~\ref{eq:H}.}
	\label{fig:annni}
\end{figure}

\begin{figure*}
	\centering
	\begin{tikzpicture}[
	connection/.style={
	->,
	line width=4pt,
	black!20!white,
	opacity=1,
	>={Triangle[length=8, width=8]}
	},
	]
	\node (dmrg) at (0,0) {\input{tikz/schema/boxes/dmrg.tex}};
	\node (brick) at (6,0) {\input{tikz/schema/boxes/brick.tex}};

	\draw[connection] (dmrg.east) -- (brick.west);
	\node[text width = 1.5cm, align=center] at ([yshift=.5cm, xshift=.9cm]dmrg.east) {\mbox{\, Circuit} \mbox{Compilation}};

	\node (hc) at (11.3,+2) {\input{tikz/schema/boxes/hc.tex}};
	\node (ae) at (11.3,-2) {\input{tikz/schema/boxes/an.tex}};

	\draw[connection]
	([yshift=.5cm]brick.east) -| ([xshift=-.45cm]hc.west) -- (hc.west);

	\draw[connection]
	([yshift=-.5cm]brick.east) -| ([xshift=-.45cm]ae.west) -- (ae.west);
\end{tikzpicture}

	\caption{\justifying Overview of the \ac{qml} pipeline. The ground state of
		the \ac{annni} model is first obtained as an \ac{mps} via \ac{dmrg}
		(left). The \ac{mps} is then compiled into a brick-layer circuit ansatz
		using the \ac{mpd} method (center). This state-preparation circuit is
		subsequently prepended to either the \ac{hc} (top right) or the \ac{qae}
		circuit (bottom right), forming two end-to-end gate-based pipelines for
		supervised and unsupervised phase classification, respectively.}
	\label{fig:scheme}
\end{figure*}

This is the simplest model combining the effect of quantum fluctuations (due to
a transverse magnetic field) and frustrated exchange interactions. Consequently,
it is a paradigm for studying competition between magnetic ordering,
frustration, and disordering effects. Competition can introduce complex effects
in spin systems~\cite{domb2000phase}, leading to the appearance of a rich phase
diagram.

Despite seeming too simplistic to describe realistic materials, this model may
reproduce important features observed experimentally in systems that can be
described by discrete models with effectively short-range competing interactions
\cite{selke1988annni}. Some of these experimental findings include Lifshitz
points, adsorbates, ferroelectrics, magnetic systems, alloys, or polytypes. On
the other hand, the so-called floating phase emerging in the model has attracted
considerable interest among experimental researchers. This critical
incommensurate phase has been observed very recently using Rydberg-atom ladder
arrays~\cite{zhang2024probing}.

Several works have already studied the structure of the phase diagram of this
model. Some of them treat the 2D classical model
\cite{selke1979monte,fisher1980infinitely,peschel1981calculation,rujan1981critical,pesch1985transfer,allen2001two,shirahata2001infinitesimal,derian2006modulation,chandra2007floating,matsubara2017domain}
and others tackle the 1D quantum model
\cite{arizmendi1991phase,sen1992numerical,rieger1996one,sen1997application,allen2001two,dutta2003gapless,beccaria2007evidence,chandra2007floating,mahyaeh2020study,fumani2021quantum,ferreira2023detecting}.
Of the latter, some deal with finite systems and others with infinite systems
\cite{nagy2011exploring}. However, in intermediate frustration and
finite-transverse field regimes, different studies report inconsistent
conclusions about the phases and phase transitions present. Two phases,
paramagnetic and floating, have been addressed
\cite{rieger1996one,chandra2007floating,nagy2011exploring}
and the former has been further divided into unmodulated and modulated ones in
some works
\cite{arizmendi1991phase,sen1992numerical,beccaria2006density,beccaria2007evidence,fumani2021quantum}.
However, the floating phase generates controversy in the literature since it was
not even found in some of the previous studies
\cite{grynberg1987alternative,sen1997application,guimaraes2002quantum,derian2006modulation}.

Building upon recent literature~\cite{Monaco_2023}, we extend the analysis of
the \ac{annni} model through \ac{qml} techniques to larger spin models, moving
from a maximum of 12 spins to 18. Beyond this numerical extension, the central
contribution is a more scalable approach to state preparation: by replacing the
\ac{vqe}-based workflow with a classical \ac{dmrg}+\ac{mps}-to-circuit pipeline,
the method can be readily deployed on real quantum hardware without per-state
optimization, and can in principle scale to system sizes well beyond those
accessible to purely quantum approaches. We investigate how this increase in
system size affects the accuracy of the \ac{qml} models and their ability to
reveal signatures of phases absent from the training data
\cite{caro2022generalization}.

On the supervised side, we also replace the \ac{qcnn} used in
Ref.~\cite{Monaco_2023} with a hybrid quantum-classical classifier (\ac{hc}),
built around a shallow, fixed-depth parametrized circuit followed by a classical
feed-forward head. This choice is motivated by a recent result showing that
\acp{qcnn} can, under fairly general conditions, be simulated efficiently on a
classical computer~\cite{bermejo2024qcnnsimulable}, which weakens the case for
running them on quantum hardware. For completeness, and to allow a direct
comparison with our earlier work, the \ac{qcnn}-based analysis is also reported
in Appendix~\ref{ap:qcnn}.

Furthermore, we subject the \ac{annni} model to an in-depth \ac{tn} analysis,
bringing the finite-size analysis to models far larger than any currently
reachable through fully quantum means, reaching up to 480 sites.

The paper is organized as follows: Section~\ref{sec:method} outlines the
state-preparation pipeline and the \ac{qml} architectures used in the study.
Section~\ref{sec:QMLanalysis} explores the analysis of the phase diagram,
utilizing both supervised and unsupervised approaches through Quantum Circuit
simulations. The structure of both analyses is depicted in
Figure~\ref{fig:scheme}. The paper concludes with a discussion of the results
and suggestions for future work in Section~\ref{sec:summary}.

\section{Methods on Quantum Machine Learning} \label{sec:method}

In the context of QML models for detecting phases in spin systems, we consider
two complementary models. The first, a hybrid quantum-classical classifier
(\ac{hc}), is a supervised learning model where the labels represent the
different phases. The second, known as \ac{qae}, follows an unsupervised
approach.

\subsection{States preparation}
\label{sec:state}

Both aforementioned \ac{qml} architectures use the ground states of the
Hamiltonian of the \ac{annni} model (Eq.~\ref{eq:H}) at different values of
$\kappa$ and $h$ as inputs. A standard approach to load these states into
quantum circuits is to use \ac{vqe}, which finds parameters $\vec{\theta}^*$ for
a given ansatz $U$ such that $U(\vec{\theta}^*)|0\rangle^{\otimes N}$
approximates the ground-state wavefunction~\cite{vqe_original}. This was the
approach adopted in Ref.~\cite{Monaco_2023}.

\subsubsection*{Why VQE does not scale}

While \ac{vqe} has been demonstrated for small systems, it presents fundamental
scalability barriers that make it impractical for the regime studied here.
First, computing gradients on quantum hardware via the parameter-shift rule
requires two circuit evaluations per parameter per training
step~\cite{vqe_original}. Since the number of parameters grows with system size,
and since the full phase diagram requires training independently at each
$(\kappa, h)$ point, the total cost scales unfavorably with both $N$ and the
number of phase diagram points. Second, in the non-simulable regime, precisely
where quantum hardware would offer an advantage, \ac{vqe} must be run entirely
on device with no classical shortcuts available. As a result, the runtime grows
with the number of phase diagram points, the number of parameters, and the
number of shots per gradient step, and quickly becomes impractical. Third,
\ac{vqe} is inherently susceptible to barren plateaus~\cite{2018NatCo...9.4812M}
and hardware noise, both of which worsen with increasing system size.

\subsubsection*{MPS-to-circuit conversion via Matrix Product Disentanglers}

In this work, we circumvent these issues by decoupling state preparation from
\ac{qml} training. Ground states are obtained classically via \ac{dmrg} as
\acp{mps}, which provide a highly accurate and scalable approach for
one-dimensional systems with short-range interactions such as the \ac{annni}
model. The \ac{mps} representation is then converted into an explicit quantum
circuit ansatz using the \ac{mpd} algorithm of Ref.~\cite{ran2020}.

The \ac{mpd} method constructs a sequence of unitary layers that
\emph{disentangle} the target \ac{mps} toward a product state using only one-
and two-qubit gates. Inverting this sequence yields a state-preparation circuit
$U_{\mathrm{MPS}}$ satisfying
\begin{equation}
	U_{\mathrm{MPS}} \,|0\rangle^{\otimes N} \approx |\psi_{\mathrm{MPS}}\rangle,
	\label{eq:mpd}
\end{equation}
where the approximation improves systematically as the number of \ac{mpd} layers
increases. The circuit $U_{\mathrm{MPS}}$ is then directly prepended to the
\ac{hc} or \ac{qae}, forming a fully gate-based, hardware-compatible pipeline
that requires no access to the full wavefunction at any stage.

The \ac{annni} model is sufficiently local, with only nearest-neighbor and
next-nearest-neighbor interactions, to justify the use of \acp{mps}. We observed
that \ac{mps} achieves high accuracy at relatively low bond dimensions, matching
exact results at bond dimension as low as $\chi=32$. Extensive testing on
systems of various sizes, revealed no noticeable decrease in accuracy with
increasing system size, and the accuracy remained consistently high along the
transition lines. The maximum bond dimension used is $\chi=64$.

For the \ac{annni} model studied here, the \ac{mpd} conversion is not exact in
general, since the \ac{mps} can have bond dimension $\chi > 2$. However, we find
that a small number of \ac{mpd} layers is sufficient to reach high fidelity,
reflecting the relatively low entanglement of the \ac{annni} ground states at
the system sizes and bond dimensions considered. Concretely, for system sizes
$N=6$, $12$, and $18$, we used $4$, $5$, and $6$ \ac{mpd} layers respectively,
yielding average fidelities of $0.9881$, $0.9672$, and $0.9514$. Unlike
\ac{vqe}, \ac{dmrg} computes the ground state for each $(\kappa, h)$ point
classically, with no gradient evaluations on hardware, and the \ac{mpd}
conversion that follows is deterministic and requires no variational
optimization. Accuracy is therefore controlled by the number of \ac{mpd} layers,
not by barren plateaus or hardware noise during
optimization~\cite{rudolph2023decomposition, dborin2022matrix}.

\begin{figure}
	\centering
	\resizebox{\linewidth}{!}{\begin{tikzpicture}
	\pgfdeclarelayer{background}
	\pgfsetlayers{background,main}

	\node (mps) at (0, 0) {\input{tikz/schema/mps4}};
	\node[anchor=west] (cir) at ([xshift=.7cm]mps.east) {\input{tikz/circuits/brick4d2}};
	\begin{pgfonlayer}{background}
		\draw[
		-{Triangle[width=10pt,length=8pt]},
		line width=6pt,
		color=gray!20
		] ([xshift=.2cm]mps.east) -- ([xshift=.1cm]cir.west);
	\end{pgfonlayer}

	\node (d1) at ([yshift=-.15cm, xshift=2.6cm]cir.south west) {$\underbrace{\rule{2.75cm}{0pt}}_{\text{Depth 1}}$};
	\node (d1) at ([yshift=-.15cm, xshift=5.5cm]cir.south west) {$\underbrace{\rule{2.75cm}{0pt}}_{\text{Depth 2}}$};
\end{tikzpicture}} \caption{\justifying
		Conversion of an \ac{mps} into a brick-layer quantum circuit via the
		\ac{mpd} method~\cite{ran2020}. The \ac{mps} (left) is faithfully
		represented by a sequence of two-qubit unitary layers (right).}
	\label{fig:conversion}
\end{figure}

\subsection{Hybrid Quantum-Classical Classifier}
\label{sec:hc}

\begin{figure*}
	\centering
	\resizebox{.9\linewidth}{!}{\definecolor{hcinputcolor}{HTML}{0096FF}
\definecolor{hcnncolor}{HTML}{4CAF50}
\newcommand{\ygate}{\gate{R_y(\cdot)}}
\newcommand{\zgate}{\gate{R_z(\cdot)}}
\newcommand{\xgate}{\gate{R_x(\cdot)}}

\tikzset{nnnode/.style={circle, draw, fill=white, minimum size=9pt, inner sep=0pt}}

\begin{tikzpicture}
	\pgfdeclarelayer{background}
	\pgfsetlayers{background,main}

	\node (circuit) {%
		\begin{quantikz}[column sep=5pt, row sep={0.7cm,between origins}]
			\gategroup[6,steps=3,style={inner xsep=1pt,fill=hcinputcolor!10}, background]{{\sc }} & \gate[6,nwires={1,2,3,4,5,6}, style={fill=hcinputcolor!20, rounded corners}]{\ket{\psi(\kappa,h)}} & \qw & \qw & \qw\gategroup[6,steps=11,style={fill=red!10, inner xsep=-2pt}, background]{{\sc Parametrized Quantum Circuit}} & \gate{R_y(\cdot)} & \ctrl{1} & \gate{R_x(\cdot)} & \qw      & \gate{R_y(\cdot)} & \ctrl{1} & \gate{R_x(\cdot)} & \qw      & \meter{} & \setwiretype{n} \\
			                                                                                      &                                                                                                    & \qw & \qw & \qw                                                                                                            & \gate{R_y(\cdot)} & \phase{} & \gate{R_x(\cdot)} & \ctrl{1} & \gate{R_y(\cdot)} & \phase{} & \gate{R_x(\cdot)} & \ctrl{1} & \meter{} & \setwiretype{n} \\
			                                                                                      &                                                                                                    & \qw & \qw & \qw                                                                                                            & \gate{R_y(\cdot)} & \ctrl{1} & \gate{R_x(\cdot)} & \phase{} & \gate{R_y(\cdot)} & \ctrl{1} & \gate{R_x(\cdot)} & \phase{} & \meter{} & \setwiretype{n} \\
			                                                                                      &                                                                                                    & \qw & \qw & \qw                                                                                                            & \gate{R_y(\cdot)} & \phase{} & \gate{R_x(\cdot)} & \ctrl{1} & \gate{R_y(\cdot)} & \phase{} & \gate{R_x(\cdot)} & \ctrl{1} & \meter{} & \setwiretype{n} \\
			                                                                                      &                                                                                                    & \qw & \qw & \qw                                                                                                            & \gate{R_y(\cdot)} & \ctrl{1} & \gate{R_x(\cdot)} & \phase{} & \gate{R_y(\cdot)} & \ctrl{1} & \gate{R_x(\cdot)} & \phase{} & \meter{} & \setwiretype{n} \\
			                                                                                      &                                                                                                    & \qw & \qw & \qw                                                                                                            & \gate{R_y(\cdot)} & \phase{} & \gate{R_x(\cdot)} & \qw      & \gate{R_y(\cdot)} & \phase{} & \gate{R_x(\cdot)} & \qw      & \meter{} & \setwiretype{n}
		\end{quantikz}
	};

	\begin{scope}[shift={($(circuit.east)+(0.5cm,-.1cm)$)}]
		\node[nnnode] (in1) at (0,1.62) {};
		\node[nnnode] (in2) at (0,0.92) {};
		\node[nnnode] (in3) at (0,0.22) {};
		\node[nnnode] (in4) at (0,-0.48) {};
		\node                (indots) at (0,-1.18) {$\vdots$};
		\node[nnnode] (in5) at (0,-1.88) {};

		\node[nnnode] (hid1) at (1.3,1.75) {};
		\node[nnnode] (hid2) at (1.3,1.25) {};
		\node[nnnode] (hid3) at (1.3,0.75) {};
		\node[nnnode] (hid4) at (1.3,0.25) {};
		\node[nnnode] (hid5) at (1.3,-0.25) {};
		\node                (hiddots) at (1.3,-0.75) {$\vdots$};
		\node[nnnode] (hid6) at (1.3,-1.25) {};
		\node[nnnode] (hid7) at (1.3,-1.75) {};

		\node[nnnode] (out1) at (2.6,0.6) {};
		\node[nnnode] (out2) at (2.6,0) {};
		\node[nnnode] (out3) at (2.6,-0.6) {};

		\foreach \i in {in1,in2,in3,in4,in5}
		\foreach \j in {hid1,hid2,hid3,hid4,hid5,hid6,hid7}
		\draw[gray, thin] (\i) -- (\j);
		\foreach \j in {hid1,hid2,hid3,hid4,hid5,hid6,hid7}
		\foreach \k in {out1,out2,out3}
		\draw[gray, thin] (\j) -- (\k);

		\node[anchor=west, font=\small, xshift=10pt] at (out1.east) {$p_1(\kappa,h)=$ Ferromagnetic};
		\node[anchor=west, font=\small, xshift=10pt] at (out2.east) {$p_2(\kappa,h)=$ Paramagnetic};
		\node[anchor=west, font=\small, xshift=10pt] at (out3.east) {$p_3(\kappa,h)=$ Antiphase};

		\coordinate (pqcbottom) at (1.3,-2.15);

		\begin{pgfonlayer}{background}
			\node[draw, thick, fill=green!20,
				fit=(in1)(in5)(hid1)(hid7)(out1)(out3)(pqcbottom), inner sep=8pt] (nnbox) {};
		\end{pgfonlayer}
		\node[anchor=south, yshift=0.0cm] at (nnbox.north) {\sc Classical Neural Network};
	\end{scope}
\end{tikzpicture}}
	\caption{\justifying Architecture of the \ac{hc} for $N=6$ qubits: input
		state preparation (blue), followed by the four-layer parametrized ansatz of
		alternating single-qubit $R_y$ and $R_x$ rotations and nearest-neighbor
		controlled-phase gates (red), a final computational-basis measurement, and a
		classical feed-forward network (green) that maps the measured features to
		the predicted phase.}
	\label{fig:hybrid}
\end{figure*}

The \ac{hc} pairs a shallow \ac{pqc} that produces a small set of measured
features with a classical feed-forward network that turns those features into a
phase prediction. The ansatz itself is a free hyperparameter of the model, and
any circuit acting on the full register can in principle be used.

Concretely, the state-preparation circuit $U_{\mathrm{MPS}}$
(Sec.~\ref{sec:state}) is followed by a brickwork ansatz made of alternating
layers of single-qubit $R_y$ and $R_x$ rotations and nearest-neighbor
controlled-phase gates. After a few such layers, the circuit is read out through
a fixed set of Pauli-Z correlators, up to a chosen correlation order: the
single-qubit expectation values $\langle Z_i\rangle$ and, when a higher order is
used, pairwise correlators $\langle Z_i Z_j \rangle$, and so on. These
expectation values form a fixed-length feature vector, which is passed through a
feed-forward network with a single hidden layer, a ReLU activation, and a final
softmax that outputs the predicted probability for each phase.

Figure~\ref{fig:hybrid} sketches the resulting circuit and processing pipeline.

\subsection{Quantum Auto Encoder}
The \ac{qae} is a quantum circuit architecture (first introduced in
\cite{PRR_anomaly_detection}) inspired by classical deep learning autoencoders.
In classical settings, autoencoders consist of an encoder that compresses input
data into a latent representation and a decoder that reconstructs the original
data. In the quantum case, the \ac{qae} performs a similar task, compressing
quantum states into a reduced subsystem.

Because quantum circuits are inherently reversible, the encoder can be inverted
to function as a decoder. For this reason, only the forward encoding part of the
circuit is trained in this framework. The goal is to disentangle a quantum state
into the tensor product of two components: a subset of qubits initialized to the
ground state $|0\rangle$ and a compressed state that retains the relevant
quantum information.

Formally, given an input state $|\psi\rangle^N$ with $N$ qubits, the \ac{qae}
applies a parameterized unitary operation $\mathcal{U}(\vec{\theta})$ such that:
\begin{equation}
	\mathcal{U}(\vec{\theta})|\psi\rangle^N = |0\rangle^{\otimes T} \otimes |\phi\rangle^{N-T},
	\label{eq:compress}
\end{equation}
where $T$ is the number of trash qubits that are mapped to the $|0\rangle$
state, and $|\phi\rangle^{N-T}$ represents the compressed quantum state
containing the essential information.

The training process minimizes the Pauli-Z expectation values of the trash
qubits, ensuring they are disentangled and collapse to $|0\rangle$. The
remaining $N-T$ qubits carry the compressed representation of the input state.

Figure~\ref{fig:as} illustrates an example of a 6-qubit \ac{qae} ($N=6$) with 3
trash qubits ($T=3$). In this case, the model disentangles 3 qubits into the
ground state while encoding the relevant quantum information into the remaining
3 qubits.

\begin{figure}
	\centering
	\resizebox{\linewidth}{!}{\input{tikz/circuits/autoencoder.tex}}
	\caption{\justifying Architecture of the \ac{qae} circuit (yellow) and input
		state preparation (blue). An example architecture for 6 qubits, with half
		designated \textit{trash} qubits positioned at the center
		\cite{Monaco_2023}.}
	\label{fig:as}
\end{figure}

\section{Quantum Machine Learning analysis}
\label{sec:QMLanalysis}
The upcoming simulations were conducted using noiseless quantum simulators,
meaning no quantum noise was incorporated. While this represents an idealized
scenario, the impact of noise is expected to be limited due to two primary
factors. First, both architectures just presented are relatively shallow, which
inherently reduces their vulnerability to noise. Second, variational quantum
models are generally regarded as being robust to moderate levels of noise, as
suggested by Ref.~\cite{giulio24}.

\subsection{Semi-Supervised approach: Hybrid Classifier}
\label{sec:hc_analysis}
\def\heightfigsqcnn{5cm} \def\heightfigsqaut{5cm}

The \ac{hc} introduced in Sec.~\ref{sec:hc} offers a supervised approach to
detecting the phase transitions of the \ac{annni} model, in the same spirit as
the \ac{qcnn} analysis of Appendix~\ref{ap:qcnn}, but built around a shallow,
fixed-depth quantum circuit whose parameters are optimized jointly with those of
the classical feed-forward head. The classical head outputs a probability
distribution over three phases: Ferromagnetic, Paramagnetic, and Antiphase.

The training procedure consists of the minimization of the following
temperature-sharpened cross-entropy loss function:
\begin{equation}
	\mathcal{L} = -\frac{1}{|S|} \sum_{(\kappa, h) \in S} \sum_{j} y_j^{\frac{1}{T}}(\kappa, h) \log\!\left(p_j(\kappa, h)\right)^{\!\frac{1}{T}},
	\label{eq:hc_loss}
\end{equation}

where $S$ is the training set, $p_j(\kappa, h)$ is the model's predicted
probability of the system at $(\kappa, h)$ being in the $j$-th phase,
$y_j(\kappa, h)$ represents the one-hot encoded labels for the three phases, and
$T$ is a temperature factor that controls the sharpness of the predicted
probability distribution.

Following the approach outlined in Ref.~\cite{Monaco_2023}, the training set $S$
consists only of the points within the \textit{integrable} part of the phase
diagram, $\{(\kappa,h) \in \{0\} \times [0,2] \} \cup \{(\kappa,h) \in [0,1]
	\times \{0\} \}$, specifically those on the two axes: $\kappa=0$ (representing
the simple transverse field case) and $h=0$ (representing the quasi-classical
model). This choice enables the analytical derivation of labels but excludes any
points associated with the floating phase, as none exist on these axes, making
the classifier entirely agnostic to the fourth phase.

The analysis was performed for different values of the number of spins $N$,
specifically 6, 12, and 18, to accurately examine any behavior or dependency
with an increasing number of spins.

Following the training process, the model is subsequently employed across the
entire phase diagram to predict the phase at each combination of $\kappa$ and
$h$, leveraging the high generalization capabilities highlighted in Ref.
\cite{caro2022generalization} for quantum models.

\begin{figure*}
	\centering
	\subcaptionbox*{}{\includegraphics[height =
			\heightfigsqcnn]{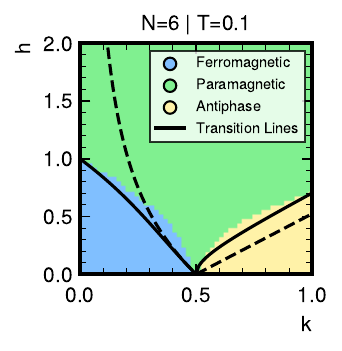} }
	\subcaptionbox*{}{\includegraphics[height =
			\heightfigsqcnn]{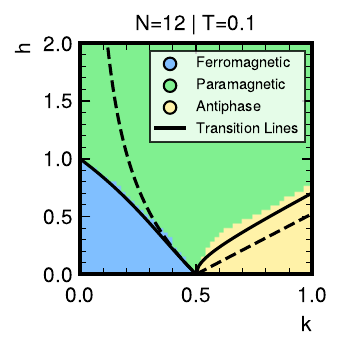} }
	\subcaptionbox*{}{\includegraphics[height =
			\heightfigsqcnn]{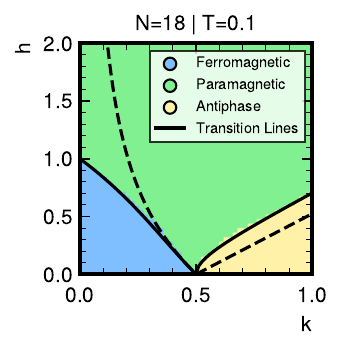} } \vskip -4ex
	\caption{\justifying Predictions of the \ac{hc} trained on the
		analytical points of the \ac{annni} Spin Model at different system
		sizes: $N=6$ (left), $N=12$ (middle), and $N=18$ (right).}
	\label{fig:hc_ana}
\end{figure*}

Figure~\ref{fig:hc_ana} displays the \ac{hc} predictions for $N=6$ (left),
$N=12$ (middle), and $N=18$ (right).

The comparison of the three images shows that increasing the number of spins
makes the classifier more precise: the predicted transition lines converge
increasingly closely to the exact ones as $N$ grows, and the model reaches a
near-perfect overall accuracy of $0.9932$ at $N=18$.

The supervised approach benefits from the high generalization capabilities of
quantum models, allowing it to perform well across the entire phase diagram.
However, identifying the sought-after floating phase is inherently out of reach
for this supervised model: since no points from that phase are present in the
training set, the classifier has no corresponding class to predict, and can only
ever assign floating-phase points to one of the three phases it was trained on.

An alternative approach that eliminates the need for labeled data is
unsupervised training, such as using the \ac{qae} architecture.

\subsection{Unsupervised approach: Quantum Autoencoder}
An alternative approach to discovering the phase landscape of a spin system
consists of employing the \ac{qae} architecture. By operating in an unsupervised
manner, it effectively bypasses the bottleneck of acquiring training labels
inherent in such a model. For this reason, the \ac{qae} architecture is a
suitable choice for identifying phases beyond the analytical ranges, such as the
floating phase.

The training process for this architecture is as follows: a single state is
selected as the training event, and the training is carried out to achieve the
compression outlined in equation \ref{eq:compress}. This involves the
minimization of the following loss function, known as \textit{compression
	score}:
\begin{equation}
	\mathcal{C} = \frac{1}{2}\sum_{j\in q_T} (1-\left<\sigma^z_j\right>),
\end{equation}
where $q_T$ refers to the selected trash qubits, constituting $N/2$ out of the
total $N$.

Hence, the \ac{qae} is trained so that it compresses the input state optimally,
leading to a low \textit{compression score}.

After training, each state of the phase diagram is compressed through the
\ac{qae}, and the corresponding cost value is assigned.

This approach does not require the model to be explicitly provided with
information about the phases. However, based on the model's output for each
state in the phase diagram, it is possible to infer both the number and the
location of each phase. Indeed, typically states belonging to the same phase as
the training event undergo optimal compression, given their inherent similarity.
Conversely, states from other phases tend to exhibit higher values in the cost
function. Notably, the compression score takes characteristic values for each
phase, allowing all phases of the spin model to be identified.

In contrast to the previous analysis~\cite{Monaco_2023}, the \ac{qae} is applied
here to larger spin models. The aim is to identify the floating phase in the
phase diagram through this model, which may have been previously obscured by the
constraints of the smaller system size.

\begin{figure*}
	\centering
	\subcaptionbox*{}{\includegraphics[height =
			\heightfigsqaut]{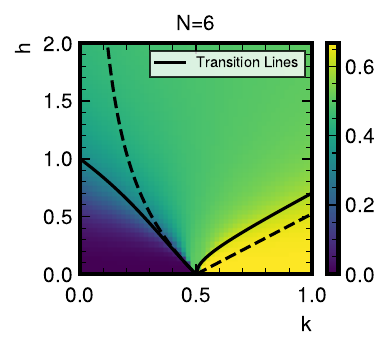} }
	\subcaptionbox*{}{\includegraphics[height =
			\heightfigsqaut]{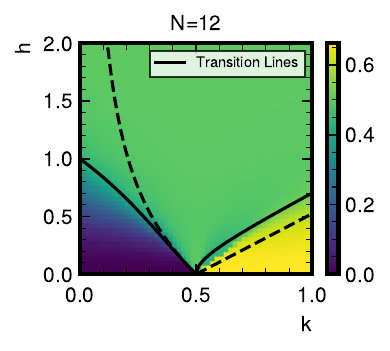} }
	\subcaptionbox*{}{\includegraphics[height =
			\heightfigsqaut]{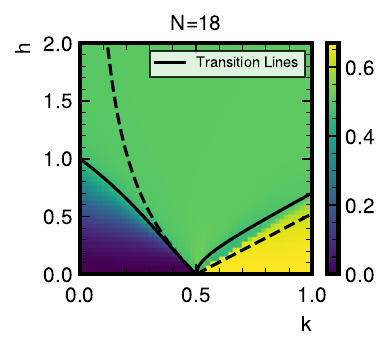} } \vskip -4ex
	\caption{ \justifying Compression Scores $\mathcal{C}$ of the
		\ac{qae} circuits trained on the $(\kappa, h) = (0, 0)$ point of the
		\ac{annni} model phase diagram at different system sizes $N$: 6
		(left), 12 (middle), and 18 (right). The scores are showcased as a
		function of the interaction strength ratio ($\kappa = -J_2/J_1$) and
		the external magnetic field ($h = B/J_1$). Lower compression scores
		indicate better disentanglement of trash qubits from others, as
		defined by eq. \ref{eq:compress}.}
	\label{fig:autoencoders}
\end{figure*}

Figure~\ref{fig:autoencoders} depicts the compression scores generated by the
\ac{qae} for $N=6$, $12$, and $18$.

All \ac{qae} models were trained to compress the point $(\kappa,h)=(0,0)$ of the
Hamiltonian, representing the case with no magnetic field and no next-nearest
neighbor interaction between spins.

Two main observations can be drawn from the figure. Firstly, the boundaries
between different phases become increasingly sharp with larger system sizes.
Secondly, up to 12 qubits (the maximum limit from the previous analysis), no
floating phase can be discerned from the compression score. However, at $N=18$,
the presence of a fourth phase becomes evident, indicated by a dark green shade
in the area associated with the floating phase.

When $N=18$, the system size is sufficiently large to identify the presence of
the floating phase. However, the model's ability to accurately reproduce the
region of the floating phase is limited by the system size, as the boundaries of
the dark green region appear imprecise and blurred. It is expected that a larger
system size would lead to a clearer definition of this region.

\section{\label{sec:summary}Summary and discussion}

In this work, we defined a \ac{qml} pipeline integrating \acp{tn} and quantum
circuits to perform the classification of different phases of the \ac{annni}
model. Starting from ground states obtained with \ac{dmrg} as \acp{mps}, we
convert these directly into quantum circuit ans\"{a}tze using the \ac{mpd}
method of Ref.~\cite{ran2020}, which encodes the \ac{mps} into layers of one-
and two-qubit gates that are prepended to the \ac{hc} or \ac{qae}. These
end-to-end gate-based circuits are compatible with real quantum hardware and
establish a concrete, circuit-level bridge between \acp{tn} and \ac{qml}. In
this configuration, we benefit from the flexibility and scalability of \acp{tn}
together with the trainability and reduced quantum resources of the \ac{hc}.

On the supervised side, we confirmed that the \ac{hc}, a shallow, fixed-depth
quantum circuit combined with a lightweight classical head, is capable of
identifying the phase transitions of the \ac{annni} model, extending the
\ac{qcnn}-based approach of Ref.~\cite{Monaco_2023} (revisited for comparison in
Appendix~\ref{ap:qcnn}) to system sizes of up to 18 sites.

Regarding classical simulability, the \ac{hc} architecture does not share the
qubit-discarding, tree-like structure that makes the \ac{qcnn} classically
tractable~\cite{bermejo2024qcnnsimulable}, and it places no restriction on the
choice of ansatz. The specific shallow, fixed-depth ansatz and low-weight
correlator readout used in our numerical experiments are, however, themselves
simulable through a light-cone argument, since bounded-depth circuits with
bounded-weight observables depend only on a limited region of the input,
regardless of system size; this is a feature of this particular choice, made for
numerical convenience. A further, and arguably stronger, route to
non-simulability would be to replace the exact expectation-value readout used
here with a measurement-sampling readout, as in the hardware-oriented model of
Ref.~\cite{hoffmann2026hybrid}.

As a supervised learning model, the \ac{hc} lacks an out-of-the-box mechanism
for training to recognize the floating phase, as no points within that phase
belong to the analytical subspace of the phase diagram.

An alternative approach, that has shown promising results, is the \ac{qae}
architecture. By employing the architecture on the \ac{annni} model at larger
system sizes than previously considered, we demonstrate that the \ac{qae} can
detect the floating phase in a fully unsupervised manner. Furthermore, the
scaling analysis presented suggests that increasing the system size further will
allow the model to accurately trace the floating phase. The possibility of
classifying or detecting this phase opens an interesting point of discussion. On
the one hand, the onsite magnetization from the \ac{mps} state is expected to
decrease with system size to accommodate the critical nature. On the other hand,
we do expect better performance of the \ac{qml} model up to a certain system
size, where the chosen local observables will vanish, i.e.\ the local order
parameters will go to zero. This could represent a turning point in the proposed
approach, where the definition of different observables should be considered,
possibly a 2-point correlation function. Other alternatives are possible, as
discussed in Ref.~\cite{khosrojerdi2024learningclassifyquantumphases}.

One of the main features shared by both architectures is the use of the
ground-state wavefunctions of the \ac{annni} model as inputs. These are obtained
classically through \ac{dmrg} and converted into quantum circuits using the
\ac{mpd} encoding of Ref.~\cite{ran2020}, which constructs the state-preparation
unitary from the \ac{mps} tensors using only one- and two-qubit gates. This
approach goes beyond earlier
works~\cite{rudolph2023decomposition,dborin2022matrix} in that the resulting
circuit is directly concatenated with the \ac{hc} or \ac{qae}, rather than being
used only for data loading in a simulator. The resulting pipeline is therefore a
hardware-native integration of \acp{tn} and \ac{qml}, enabling faster and more
accurate state preparation while preserving the scalability of \ac{dmrg}.

Another possibility for lifting the training directly onto hardware at scale is
the route of simulable losses paired with quantum-hard outputs e.g.\ with
\ac{iqp} sense~\cite{bremner2011classical}, as recently demonstrated for
generative models~\cite{recio2025trainclassical}. This keeps the full training
pipeline classical, while still requiring a quantum computer to produce the
outputs.

The \ac{tn} analysis (Appendix~\ref{sec:TNanalysis}) provides numerical evidence
of changes in the properties of the ground state, allowing us to characterize
the phase diagram. By using the recent techniques proposed in
\cite{chepiga2020floating,chepiga2021lifshitz,chepiga2023eight}, we have
identified the phase transitions appearing in the model by fixing the value of
the transverse magnetic field. We have found agreement with previous \ac{dmrg}
results~\cite{beccaria2007evidence} and perturbative analysis
\cite{suzuki2012quantum}.

However, although \ac{tn} analysis provides more accurate results as larger
system sizes can be achieved, the amount of entanglement in the system could
limit the study of more complex models. On the other hand, quantum computers
promise to simulate states of high entanglement more effectively than classical
ones. Overall, the study of phase diagrams through QML techniques seems to be a
promising avenue.

\begin{acknowledgments}
	M.C. acknowledges Natalia Chepiga for useful discussions regarding the
	nature of the \ac{kt} transition and the effects of boundary conditions. M.C
	is partly funded by the Deutsche Forschungsgemeinschaft (DFG, German
	Research Foundation) under Germany's Excellence Strategy
	--EXC-2111--390814868 and by Research Unit FOR 5522 (grant nr. 499180199).

	M.G. and S.V. are supported by CERN through the CERN Quantum Technology
	Initiative.

	E.R. acknowledges support from the BasQ strategy of the Department of
	Science, Universities, and Innovation of the Basque Government. E.R. is
	supported by the grant PID2021-126273NB-I00 funded by MCIN/AEI/
	10.13039/501100011033 and by ``ERDF A way of making Europe" and the Basque
	Government through Grant No. IT1470-22. This work was supported by the EU
	via QuantERA project T-NiSQ grant PCI2022-132984 funded by
	MCIN/AEI/10.13039/501100011033 and by the European Union
	``NextGenerationEU''/PRTR. This work has been financially supported by the
	Ministry of Economic Affairs and Digital Transformation of the Spanish
	Government through the QUANTUM ENIA project called – Quantum Spain project,
	and by the European Union through the Recovery, Transformation, and
	Resilience Plan – NextGenerationEU within the framework of the Digital Spain
	2026 Agenda.

	L.T. acknowledges support from the Proyecto Sin\'ergico CAM 2020
	Y2020/TCS-6545 (NanoQuCo-CM), Ayudas Intramurales CSIC 202250|199, the CSIC
	Research Platform on Quantum Technologies PTI-001, and from Spanish projects
	PID2021-127968NB-I00 and TED2021-130552B-C22 funded by
	MCIN/AEI/10.13039/501100011033/FEDER, UE and MCIN/AEI/10.13039/501100011033,
	respectively.

	This work has been financially supported by the Ministry of Economic Affairs
	and Digital Transformation of the Spanish Government through the QUANTUM
	ENIA project called – Quantum Spain project, and by the European Union
	through the Recovery, Transformation, and Resilience Plan – NextGenerationEU
	within the framework of the Digital Spain 2026 Agenda.

	This work has been partially funded by the Eric \& Wendy Schmidt Fund for
	Strategic Innovation through the CERN Next Generation Triggers project under
	grant agreement number SIF-2023-004.
\end{acknowledgments}

\appendix

\section{Quantum Convolutional Neural Network approach}
\label{ap:qcnn}

For completeness, and to allow a direct comparison with our earlier
work~\cite{Monaco_2023}, we also analyze the \ac{annni} phase diagram using the
\ac{qcnn} architecture originally proposed in Ref.~\cite{Cong_19} for the
detection of phase transitions. A recent result, however, shows that this
specific type of ansatz, one that progressively discards qubits through pooling
down to a small final measured register, can be simulated efficiently on a
classical computer under fairly general
conditions~\cite{bermejo2024qcnnsimulable}, which is why we adopt the \ac{hc}
architecture of Sec.~\ref{sec:hc} as the main supervised approach used
throughout this work.

\begin{figure}
	\centering
	\resizebox{\linewidth}{!}{\input{tikz/circuits/qcnn.tex}}
	\caption{\justifying Architecture of the \ac{qcnn}: input state preparation
		(blue), then alternating Convolution layers (green) and Pooling layers
		(red), and a Fully Connected layer (dark green) at the
		end~\cite{Monaco_2023}.}
	\label{fig:qcnn}
\end{figure}

\acp{qcnn} are a category of quantum circuits inspired by the widely popular
classical counterpart, \acp{cnn}, aiming to learn representations from input
data by exploiting local properties, in this case the interactions between
neighboring spins. As shown in Fig.~\ref{fig:qcnn}, the architecture is composed
of three primary components: Convolution layers, which apply alternating
unitaries to pairs of neighboring spins; Pooling layers, which measure half of
the qubits and apply a different rotation to the remaining ones depending on the
measurement outcome; and a Fully Connected layer, applied once at the end to the
surviving qubits. This structure includes a local final
measurement~\cite{cerezo2021cost} and a number of parametrized gates that scales
logarithmically with the number of qubits, both of which help suppress barren
plateaus~\cite{BP_QCNN}. Recent advancements have also significantly optimized
the implementation of this architecture on real
hardware~\cite{hua2023exploiting, chinzei2023splitting}.

As with the \ac{hc}, the input state is prepared with the \ac{mpd}-converted
circuit $U_{\mathrm{MPS}}$ of Sec.~\ref{sec:state}. Here, the labels are encoded
directly in a 2-qubit output register:\\[.01cm]

\quad \begin{tabular}{llll} $\bullet$ & Ferromagnetic & : & $|00\rangle$
               \\[.1cm]
               $\bullet$        & Paramagnetic  & : & $|01\rangle$ \\[.1cm]
               $\bullet$        & Antiphase     & : & $|10\rangle$ \\[.1cm]
               $\bullet$        & Trash Class   & : & $|11\rangle$ \\[.1cm]
\end{tabular}\\

The model is trained by minimizing the same temperature-sharpened cross-entropy
loss as the \ac{hc} (Eq.~\ref{eq:hc_loss}), again using only the points on the
analytically solvable axes $\kappa=0$ and $h=0$, following the semi-supervised
strategy of Sec.~\ref{sec:hc_analysis}. As in that case, no floating phase
samples are present in the training set, but here the fourth, otherwise unused,
output state $|11\rangle$ is available as a Trash Class.

\begin{figure*}
	\centering
	\subcaptionbox*{}{\includegraphics[height =
			\heightfigsqcnn]{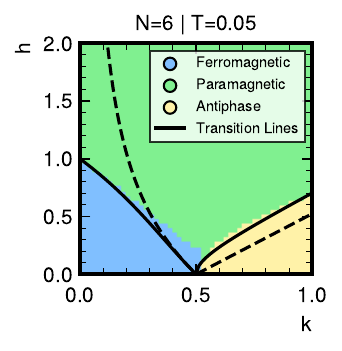} }
	\subcaptionbox*{}{\includegraphics[height =
			\heightfigsqcnn]{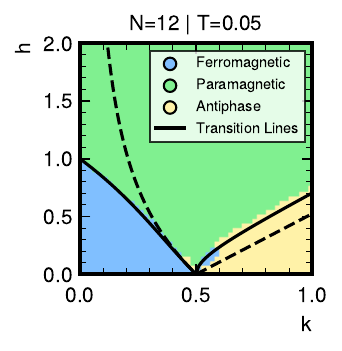} }
	\subcaptionbox*{}{\includegraphics[height =
			\heightfigsqcnn]{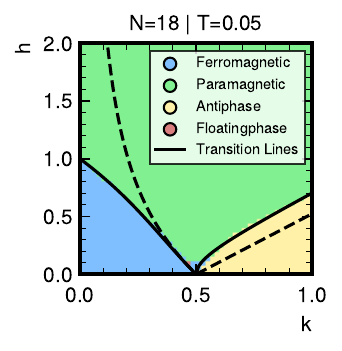} } \vskip -4ex
	\caption{\justifying Predictions of the \acp{qcnn} trained on the
		analytical points of the \ac{annni} Spin Model at different system
		sizes: $N=6$ (left), $N=12$ (middle), and $N=18$ (right). Colors
		represent Ferromagnetic (light blue), Paramagnetic (light green),
		Antiphase (light yellow), and Trash Class (light red) as a function
		of the external magnetic field ($h = B/J_1$) and interaction
		strength ratio ($\kappa=-J_2/J_1$) (refer to eq. \ref{eq:H}).}
	\label{fig:qcnn_ana}
\end{figure*}

The analysis was carried out for $N=6$, $12$, and $18$ spins, and the trained
model was then evaluated across the full phase diagram, again relying on the
generalization capabilities of quantum models~\cite{caro2022generalization}.
Figure~\ref{fig:qcnn_ana} shows the resulting predictions.

\begin{figure}
	\centering
	\includegraphics[height = \heightfigsqcnn]{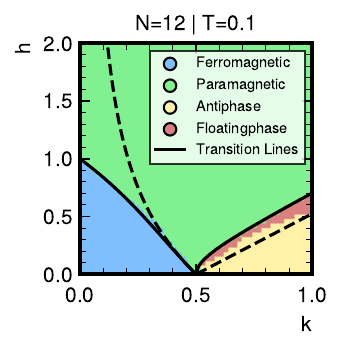}
	\caption{\justifying Example of a \ac{qcnn} prediction for $N=12$ in which
		part of the floating phase region is classified as the Trash Class
		(light red). Colors follow the same convention as in
		Fig.~\ref{fig:qcnn_ana}.}
	\label{fig:qcnn_trash}
\end{figure}

As for the \ac{hc}, the accuracy of the ferromagnetic-paramagnetic transition
improves with system size, while the predicted antiphase-paramagnetic transition
shifts from the \linefp line toward the \linefa line, consistent with the
growing presence of the floating phase at larger $N$. In rare training
instances, we additionally observed a qualitatively different behavior: a
non-negligible portion of the floating phase region was instead classified as
the Trash Class, the fourth output state on which the model was never trained.
An example for $N=12$ is shown in Fig.~\ref{fig:qcnn_trash}. The exact
conditions triggering this behavior remain unclear and appear to depend on the
random initialization of the circuit parameters; it is not entirely surprising,
though, since the floating phase, being absent from the training data, has no
assigned target class and can therefore be captured by the otherwise unused
Trash Class.

\section{Tensor network overview}
\acp{tn}
\cite{orus2014practical,schuch2013condensed,eisert2013entanglement,augusiak2012many,verstraete2008matrix,cirac2009renormalization,biamonte2017tensor,montangero2018introduction,silvi2019tensor,montangero2022loop,ran2020,banuls2023}
serve as versatile mathematical constructs designed for the representation and
manipulation of high-dimensional data, particularly in the realms of quantum
physics and \ac{ml}
\cite{stoudenmire2018learning,liu2019machine,han2018unsupervised}. \acp{tn}
provide a generalization of the original ideas beyond the \ac{dmrg}
\cite{white1992density, white1993density, ostlund1995thermodynamic,
	verstraete2004density, daley2004time, schollwock2005density,
	schollwock2011density}  which to date stands as the best algorithm to simulate
one-dimensional quantum many-body systems. A generic \ac{tn} is built to provide
an efficient encoding and manipulation of large data structures, by
appropriately compressing them into small chunks according to specific patterns
of correlations. Such small chunks become the individual tensors in the network.
\acp{tn} have witnessed widespread applications across diverse fields, including
quantum information theory, condensed matter physics at and out of equilibrium,
and deep learning~\cite{levine2019quantum,qin2022,frias-perez2023}.

\acp{mps} \cite{fannes1992finitely,klumper1991equivalence,klumper1993matrix} are
the relevant \acp{tn} for \ac{dmrg} and represent a particular class of \acp{tn}
predominantly used for depicting quantum states in one-dimensional systems.
Within the \ac{mps} framework, a quantum state is built as a product of
matrices, each corresponding to a specific site within the system. The
entanglement structure of the state is succinctly encapsulated through these
matrices, providing a concise representation of many-body quantum states. Having
proven themselves a powerful tool, \acp{mps} have been instrumental in studying
the properties of one-dimensional quantum systems
\cite{hastings2006solving,hastings2007area}, involving aspects like ground state
properties and dynamic behaviors.

In the context of \acp{mps}, \ac{dmrg} \cite{white1992density, white1993density,
	ostlund1995thermodynamic, verstraete2004density, daley2004time,
	schollwock2005density, schollwock2011density} is an iterative numerical
algorithm that allows one to obtain the best \ac{mps} for encoding the
ground state of quantum systems in one and two dimensions
\cite{tagliacozzo2009,yan2011spin,stoudenmire2012studying,zheng2017stripe,jiang2019superconductivity,qin2020absence}.

Originally formulated for gapped Hamiltonians, \ac{dmrg} can also be used to
describe critical systems by leveraging an appropriate finite-size scaling
procedure
\cite{cardy1984,holzhey1994geometric,vidal2003entanglement,latorre2003ground,calabrese2004entanglement}
or by directly working in the thermodynamic limit
\cite{nishino_1996,tagliacozzo_2008,pollmann2009}. Beyond its origins in
condensed matter physics, it has evolved into a standard tool, finding
applications in diverse domains, including quantum chemistry, high-energy
physics, and \ac{ml}.

As a result, \ac{dmrg} has emerged as the \textit{de facto} standard for
analyzing phase diagrams of one-dimensional systems. Here, to make contact with
\ac{qml} results we characterize the phase diagram on a finite chain, and thus
make use of \ac{fss} techniques~\cite{henkel1999conformal, fisher1972scaling,
	binder1992monte} which allows us to make predictions about the thermodynamic
limit by studying local order parameters of finite systems. In addition to this,
some insights on one-dimensional quantum field theory
\cite{giamarchi2003quantum} are necessary to understand some of the phases and
their limits. Numerical simulations have been carried out with TeNPy's two-site
\ac{dmrg} \cite{hauschild2018efficient}.

\section{Tensor Network analysis} \label{sec:TNanalysis}

\begin{figure}
	\includegraphics[width=\linewidth]{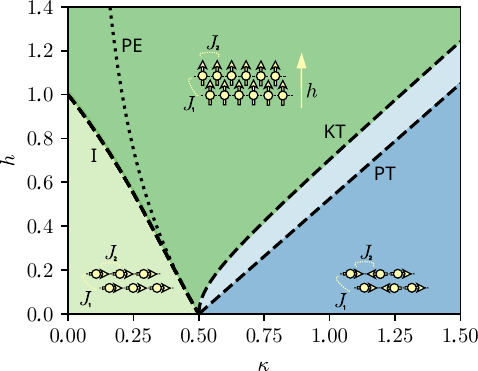}
	\caption{\label{fig:Phase_diagram} \justifying Phase diagram of the quantum
		one-dimensional \ac{annni} model of Eq.~\ref{eq:H}. The dotted line is
		the exactly-solvable Peschel-Emery (PE) line $h=1/4\kappa - \kappa$
		\cite{peschel1981calculation}. Dashed lines represent Ising (I),
		Kosterlitz-Thouless (KT), and Pokrovsky-Talapov (PT) phase transitions.
		The spin chain can be seen as a ladder of two spin chains as sketched in
		the cartoon spin configurations. }
\end{figure}

From earlier studies including simulations of both finite
\cite{beccaria2007evidence} and infinite~\cite{nagy2011exploring,
	verresen2019stable} systems and the study of a spin chain that can be mapped
onto our model using a non-local transformation
\cite{verresen2019stable,chepiga2023eight}, we expect a phase diagram like the
one in Fig.~\ref{fig:Phase_diagram}, with four different phases (the
ferromagnetic phase, the paramagnetic phase, the floating phase and the
antiphase) and a disordered line~\cite{peschel1981calculation}. All phases are
gapped except the gapless floating phase. A system is known to be gapless if
there are excitations at arbitrarily low energies in the infinite lattice limit
(and gapped if not)~\cite{sachdev1999quantum}. In this work, as we simulate
finite systems, there is necessarily a nonzero energy separating the ground
state and the first excited state. However, this energy spacing can either
remain finite or approach zero in the thermodynamic limit.

To observe the transitions and the properties of the different phases, we fix
the value of the transverse field, $h$, and vary the frustration parameter,
$\kappa$. When the system goes through a phase transition (i.e., at a critical
point), a qualitative change in correlations of the ground state occurs. A
measure of the range over which these correlations approach zero is called the
correlation length, $\xi$. In gapped phases, $\xi$ is finite and the correlation
functions of local operators typically exhibit exponential decay with distance,
\begin{equation}
	C_{i,j}^{zz}=\langle\sigma^{z}_{i}\sigma^{z}_{j}\rangle-\langle\sigma^{z}_{i}\rangle\langle\sigma^{z}_{j}\rangle \propto e^{-|i-j|/\xi}.
\end{equation}
In gapless phases, this quantity can become infinite and the correlation
functions follow a power-law behavior with distance.
\begin{equation}
	C_{i,j}^{zz}\propto \frac{1}{|i-j|^\eta}.
\end{equation}
When the system is approaching a critical point, the correlation length diverges
and how it does so allows us to know the nature of the transition. Note that the
correlation length extracted from different operators may differ in value, but
all are proportional and scale in the same way with system size.

\begin{figure}
	\includegraphics{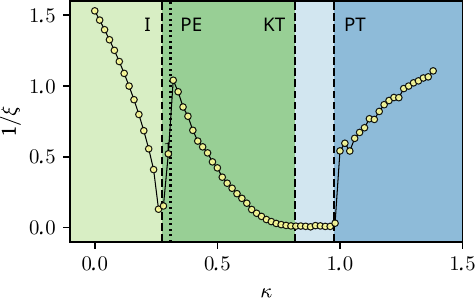}
	\caption{\label{fig:Correlation} \justifying Inverse of the correlation
		length computed from $C_{i,j}^{zz}$ as a function of $\kappa$ along the
		horizontal cut at $h=0.5$ obtained on a finite-size system with $N=240$
		sites with open boundary conditions. Colors indicate different phases:
		the ferromagnetic phase (light green), the paramagnetic phase (dark
		green), the floating phase (light blue), and the antiphase (dark blue)
		from left to right. Phase transitions according to previous results
		\cite{peschel1981calculation,suzuki2012quantum,beccaria2007evidence} are
		also shown by dashed lines: Ising (I) transition, Kosterlitz-Thouless
		(KT) transition, and Pokrovsky-Talapov (PT) transition. The dotted line
		is the exactly solvable Peschel-Emery (PE) line
		\cite{peschel1981calculation}.}
\end{figure}

In Fig.~\ref{fig:Correlation}, the inverse of the correlation length along the
horizontal cut at $h=0.5$ is shown, providing qualitative insights into the
nature of the phase transitions. At low values of frustration, the inverse of
the correlation length vanishes linearly in agreement with the Ising (I)
critical exponent $\nu=1$. After this, it reaches its maximum at a sharp kink
that corresponds to the disorder point
\cite{peschel1981calculation,pesch1985transfer}. Beyond this point, it decreases
quickly, in agreement with the exponential divergence of the correlation length
typical for a Kosterlitz-Thouless (KT) transition
\cite{shirakura2014kosterlitz}. Approaching the phase transition between the
critical phase and the antiphase from the latter, the inverse of the correlation
length vanishes with a critical exponent smaller than one, in agreement with the
Pokrovsky-Talapov (PT) critical exponent $\nu=1/2$ \cite{pokrovsky1979ground}.
Details can be found in Appendix \ref{ap: Correlation_length}.

In the following sections we present results obtained with state-of-the-art
techniques on system sizes beyond previous studies. Although the TN analysis is
not the main focus of this work, we highlight that it represents an independent
contribution to the computational characterization of the model.

\subsection{Ferromagnetic Phase and Paramagnetic Phase}

We now examine more closely the phase transition between the ferromagnetic and
paramagnetic phases.

In the ferromagnetic phase, the next neighbor interactions dominate and all the
spins are aligned in the x direction,
$\ket{\rightarrow\rightarrow\cdots\rightarrow}$ or
$\ket{\leftarrow\leftarrow\cdots\leftarrow}$ (note that
$\sigma^{x}_{j}\ket{\leftrightarrows}=\pm\ket{\leftrightarrows}$), leading to a
uniform magnetization. In the paramagnetic phase, the transverse magnetic field
dominates and aligns the spins in the $z$ direction,
$\ket{\uparrow\uparrow\cdots\uparrow}$ (note that
$\sigma^{z}_{j}\ket{\uparrow}=\ket{\uparrow}$). As a result, the magnetization
along the $x$ direction is a relevant order parameter to characterize the
quantum phase transition,
\begin{equation}
	M^{x}=\frac{1}{N}\sum_{j=1}^{N}\sigma^{x}_{j}.
	\label{eq:magnetization}
\end{equation}

At low values of frustration, the magnetization is expected to be non-zero,
indicating an ordered phase. As the frustration increases, there is a critical
value where the magnetization undergoes a continuous change, signaling a quantum
phase transition to a disordered phase with the magnetization approaching zero.
This change is rounded when the system size $N$ is finite
\cite{henkel1999conformal}. Consequently, the critical frustration value at
which the phase transition occurs may shift slightly for finite systems compared
to the thermodynamic limit. To make predictions of this limit by working with
finite systems, we make use of Finite-Size Scaling (FSS)
\cite{fisher1972scaling,henkel1999conformal}. Within this framework, Binder's
cumulant can be used to locate the precise transition point
\cite{binder1992monte}

\begin{equation}
	U_{_{N}}=1-\frac{\langle(M^{^{x}})^{^{4}}\rangle_{_{N}}}{3\langle(M^{^{x}})^{^{2}}\rangle^{^{2}}_{_{N}}}.
	\label{eq:Binder}
\end{equation}

Since this ratio is built to be size-independent at the critical point, by
plotting it as a function of the parameter of the model for different system
sizes, one can identify the critical point as the crossing of the lines.

\begin{figure}[t]
	\hspace*{-0.5cm}
	\includegraphics{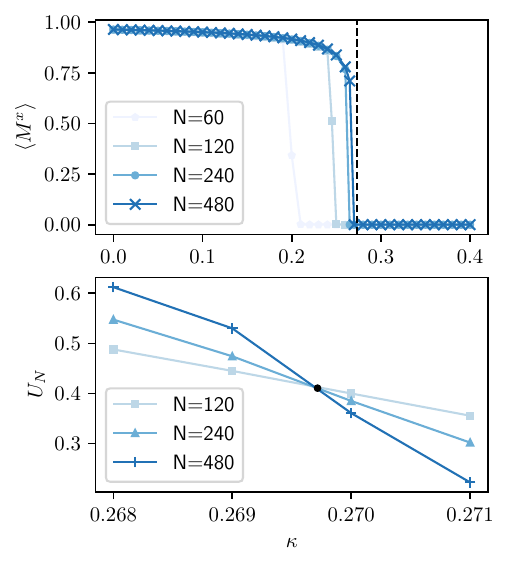}
	\caption{\label{fig:mx} \justifying (Up) Magnetization along $x$ (see Eq.
		\ref{eq:magnetization}) obtained on four finite-size systems
		($N=60,120,240,480$) with open boundary conditions along the horizontal
		cut at $h=0.5$. As expected, magnetization in the ferromagnetic phase is
		non-zero and in the paramagnetic phase is zero. Note that the phase
		transition approaches the perturbative analysis prediction as the length
		of the chain increases. (Down) Binder's cumulant (see Eq.
		\ref{eq:Binder}) for $\kappa$-values in the surroundings of the critical
		point extracted from Eq.~\ref{eq:I_transition}.}
	\label{fig:magnetization}
\end{figure}

The phase boundary between the ferromagnetic and the paramagnetic phases can be
obtained by using perturbation theory for $\kappa<0.5$, splitting the
Hamiltonian in a classical Ising part and a small transverse fluctuation, an
approximate expression for the critical line is obtained in
\cite{chakrabarti2008quantum} and reads

\begin{equation}
	h_{_{I}}\approx\frac{1-\kappa}{\kappa}\left(1-\sqrt{\frac{1-3\kappa+4\kappa^{2}}{1-\kappa}}\right),
	\label{eq:I_transition}
\end{equation}
and corresponds to an Ising-like transition~\cite{suzuki2012quantum}.

Fig.~\ref{fig:magnetization} (upper panel) shows the behavior of the local order
parameter $\langle M^{x}\rangle$, which allows us to differentiate both the
ferromagnetic and the paramagnetic phases. Finite-size effects appear and it can
be seen how the change in magnetization approaches the critical point and
becomes steeper as the system size increases. By computing Binder's cumulant
through Eq.~\ref{eq:Binder}, we get $\kappa_{_{I}}=0.270$, see Fig.
\ref{fig:magnetization} (lower panel). This implies a relative error of $1.39\%$
to the perturbative analysis calculations.

\subsection{Floating Phase}

\subsubsection{Luttinger Liquid and Friedel Oscillations}

\begin{figure}
	\includegraphics[width=\linewidth]{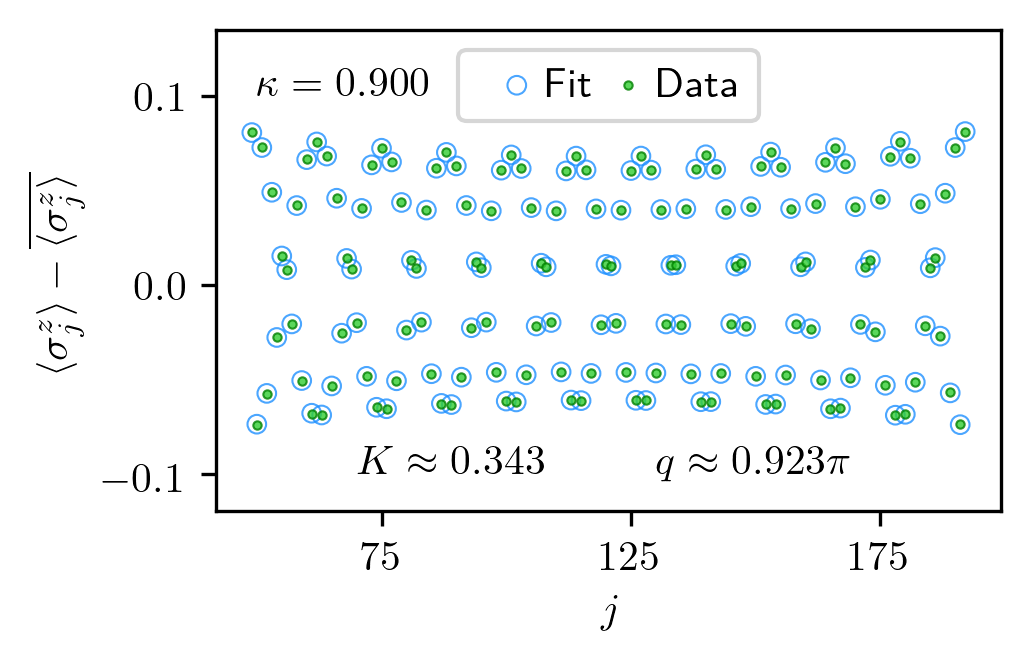}
	\caption{\label{fig:Friedel_oscillation_patterns} \justifying Friedel
		oscillation pattern inside the floating phase for $h=0.5$ and
		$\kappa=0.900$ obtained on a finite-size system of $N=240$ sites with
		open boundary conditions. Green points are \ac{dmrg} data and blue
		circles are the result of the fit with Eq.~\ref{eq:Friedel} (note the
		very accurate agreement). The uniform part of the spin-density
		$\overline{\langle\sigma^{z}_{j}\rangle}$ has been subtracted (see
		Appendix \ref{ap: Luttinger_parameters}), and the fitting window is
		restricted to the range $j\in[48,191]$ to avoid edge effects.}
\end{figure}

In 1D quantum physics, the floating phase~\cite{bak1982commensurate,
	pokrovsky1979ground, huse1984commensurate} is described by a Luttinger
liquid with algebraic incommensurate correlations
\cite{giamarchi2003quantum}. Its properties are described by a bosonic
conformal field theory and the decay of all the correlation functions is
controlled by the parameter $K$, often referred to as the Luttinger liquid
parameter~\cite{giamarchi2003quantum, haldane1981luttinger}. This Luttinger
liquid phase is stable against superconducting perturbations and spontaneous
translation symmetry breaking due to an emergent $U(1)$ symmetry when the
Luttinger exponent lies within the interval $1/4<K<1/2$
\cite{verresen2019stable}. Furthermore, open boundary conditions act as an
impurity and lead to Friedel oscillations in spin density. According to
\ac{bcft}, the profile takes the following form~\cite{cardy1991bulk,
	white2002friedel, fabrizio1995interacting}:

\begin{equation}
	\langle \widetilde{\sigma^z_{j}}\rangle\propto\frac{\cos(qj+\alpha)}{\left[(N/\pi)\sin(\pi j/N)\right]^{K}},
	\label{eq:Friedel}
\end{equation}
where $K$ is the Luttinger exponent, $q$ is the incommensurate wave-vector,
$\alpha$ is a phase shift and $\langle \widetilde{\sigma^z_{j}}\rangle$ is a
slowly decaying alternating term in the spin-density. By fitting Friedel
oscillations we can get an accurate estimate of both the Luttinger parameter $K$
and the incommensurate wave-vector $q$ \cite{chepiga2021lifshitz}. One example
of such fits is shown in Fig.~\ref{fig:Friedel_oscillation_patterns}, where the
uniform part of the spin-density, $\overline{\langle\sigma^{z}_{j}\rangle}$, has
been subtracted. Details can be found in Appendix \ref{ap:
	Luttinger_parameters}.

\subsubsection{Kosterlitz-Thouless transition}

\begin{figure}
	\includegraphics[width=\linewidth]{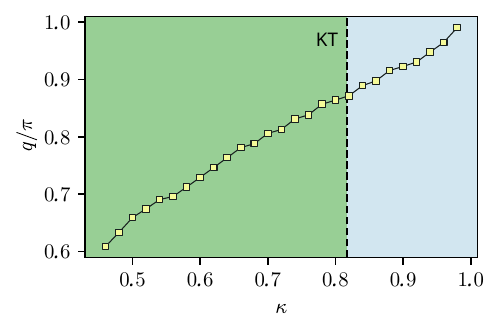}
	\caption{\label{fig:Incommensurate} \justifying Incommensurate wave-vector
		$q$ obtained on a finite-size system with $N=240$ sites with open
		boundary conditions along the horizontal cut at $h=0.5$. As it is shown,
		incommensurability doesn't vanish in the paramagnetic phase. Therefore,
		the KT transition is an incommensurate-incommensurate phase transition.}
\end{figure}

\begin{figure}
	\includegraphics{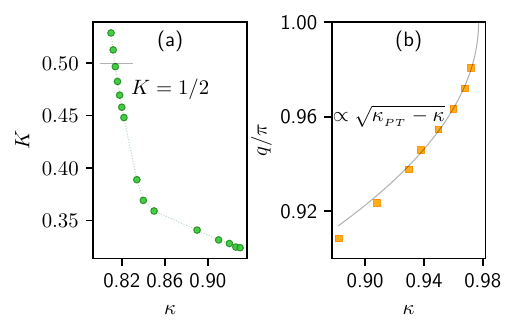}
	\caption{\label{fig:Parameters} \justifying (a) Luttinger liquid exponent
		$K$ and (b) incommensurate wave-vector $q$ obtained on a finite-size
		system of $N=240$ sites with open boundary conditions along the
		horizontal cut at $h=0.5$. The KT transition appears when $K=1/2$ and
		the PT transition arises when incommensurability vanishes. The solid
		line in (b) is a fit assuming the PT critical exponent $\nu=1/2$.}
\end{figure}

A KT transition appears between the paramagnetic phase and the floating phase.
In this model, it is an incommensurate-incommensurate transition, as we see in
Fig.~\ref{fig:Incommensurate}. As mentioned in
\cite{verresen2019stable,chepiga2023eight}, the critical point is associated
with the Luttinger liquid exponent taking the value $K=1/2$. Then, we can locate
the transition by extracting the Luttinger liquid exponent and finding the value
corresponding to $K=1/2$.

Numerical simulations involving different techniques suggest that the \ac{kt}
critical line is given by Ref.~\cite{beccaria2007evidence} as

\begin{equation}
	h_{_{KT}}(\kappa)\cong 1.05\sqrt{\left(\kappa-\frac{1}{2}\right)\left(\kappa-0.1\right)}.
\end{equation}

Fig.~\ref{fig:Parameters}(a) shows the $K$ exponent in terms of frustration. The
value we get for the critical point is $\kappa_{_{KT}}=0.814$. This means that
our result presents a relative error of $0.36\%$ concerning previous
calculations.

\subsubsection{Pokrovsky-Talapov transition}

The transition between the floating phase and the antiphase (discussed below) is
incommensurate-commensurate in the PT
\cite{dzhaparidze1978magnetic,pokrovsky1979ground} universality class. Then, we
can fit the incommensurate wave-vector $q$ as $q/\pi\propto
	(\kappa_{_{PT}}-\kappa)^{\nu}$, where $\nu$ takes the value of $\nu=1/2$ at the
\ac{pt} transition. Close to the critical point a square root-like behavior
arises and we can extract a numerical value for the critical point with high
accuracy. We obtain $\kappa_{_{PT}}=0.977$ with a root mean squared error of
$\approx 10^{-3}$.

Also from Ref.~\cite{beccaria2007evidence}, we have an expression for this
critical line given by
\begin{equation}
	h_{_{PT}}(\kappa)\cong 1.05\left(\kappa-\frac{1}{2}\right),\\
\end{equation}

from which we estimate a relative error of $0.10\%$.

\subsubsection{Entanglement entropy}
According to \ac{cft}, the entanglement entropy at critical points in a
finite-size chain with open boundary conditions scales with the block size $n$
as
\cite{affleck1991universal,holzhey1994geometric,vidal2003entanglement,calabrese2004entanglement,calabrese2009entanglement}

\begin{equation}
	S_N(n)=\frac{c}{6}\ln d_N(n)+s_1+\ln g,
	\label{eq:Calabrese-Cardy}
\end{equation}
where $d_N(n)=\frac{2N}{\pi}\sin\left(\frac{\pi n}{N}\right)$ is the conformal
distance, and $s_1$ and $\ln g$ are non-universal constants.

\begin{figure}
	\includegraphics[width=\linewidth]{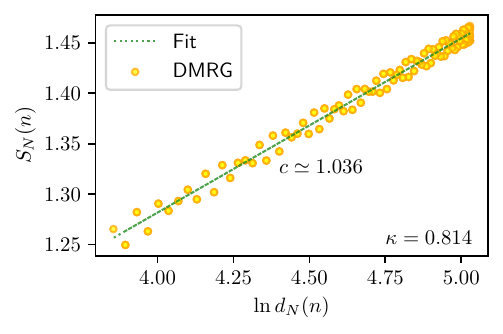}
	\caption{\label{fig:Central_charge} \justifying Entanglement entropy in
		terms of the conformal distance $d_N(n)$ at $h=0.5$ and $\kappa=0.814$
		and a finite-size system with $N=240$ sites for open boundary
		conditions. The obtained value for the central charge in the critical
		point agrees within $3.6\%$ with the \ac{cft} prediction $c=1$ for a
		Luttinger liquid.}
\end{figure}

We use Eq.~\ref{eq:Calabrese-Cardy} to extract the central charge numerically.
Close to the \ac{kt} critical line, the central charge can be extracted with
sufficient accuracy even from relatively small chains, see Fig.
\ref{fig:Central_charge} \cite{chepiga2020floating}.

The obtained value for the central charge in the critical point agrees within
$3.6\%$ for $N=240$ with the \ac{cft} prediction $c=1$ for a Luttinger liquid.
This value of the central charge could be understood if we see the model as two
quantum Ising chains coupled by a ``zig-zag'' interaction with strength $J_1$
\cite{allen2001two}. Inside the floating phase and in the large $\kappa$ limit,
the model is conformally invariant with central charge $c=1/2+1/2=1$, i.e.\ two
critical Ising models.

\subsection{Antiphase}

In the antiphase, the antiferromagnetic next-nearest neighbor interactions
dominate and the ground state is
$\ket{\rightarrow\rightarrow\leftarrow\leftarrow\ldots}$ or
$\ket{\leftarrow\leftarrow\rightarrow\rightarrow\ldots}$. From the relative
position of spins, it is easy to see that a convenient local order parameter to
identify this phase is the so-called staggered magnetization, defined as

\begin{equation}
	S^{x}=\frac{1}{N}\left(\sigma_{1}^{x}+\sigma_{2}^{x}-\sigma_{3}^{x}-\sigma_{4}^{x}+\ldots\right).
\end{equation}

However, the critical points extracted from the finite-size scaling analysis are
not as accurate as the ones extracted from the incommensurate wave vector
vanishing in the \ac{pt} transition.

\section{Extraction of the Correlation Length}\label{ap: Correlation_length}

\begin{figure}[h]
	\includegraphics{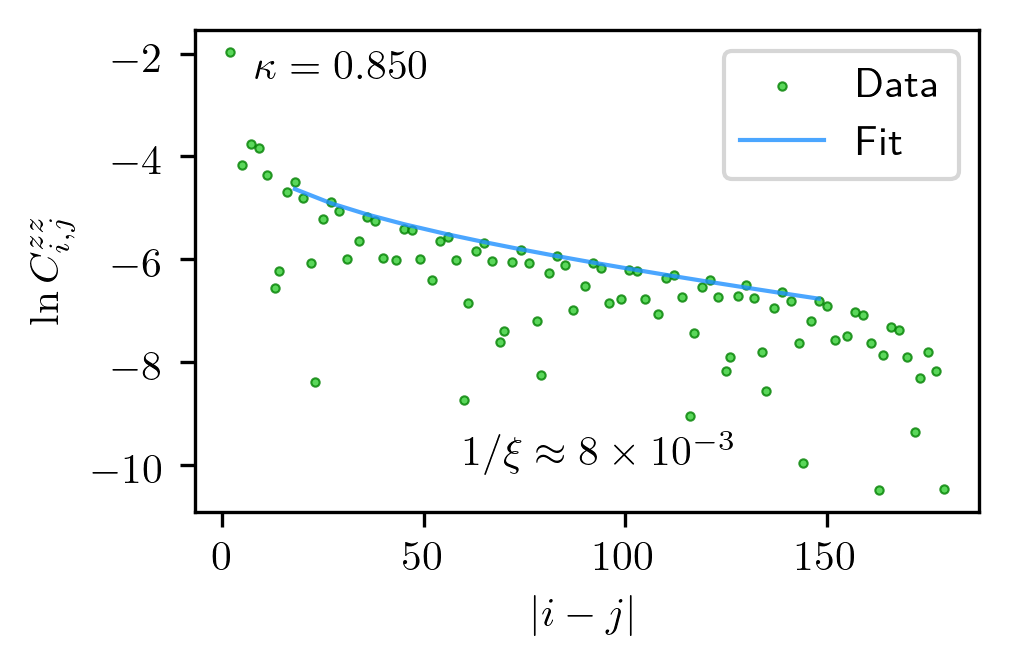}
	\caption{\justifying Extraction of the correlation length $\xi$ from
		correlations in the $z$ direction. These transverse correlations are
		obtained for $h=0.5$ inside the floating phase on a finite-size system
		with $N=240$ and open boundary conditions. The green points are
		\ac{dmrg} data and the blue line is the result of the fit with Eq.
		\ref{eq:OZ_equation}. Oscillations have been removed to fit the main
		slope of the decay.}
	\label{fig:xi_fitting}
\end{figure}

Despite the different behavior of the correlation length in gapped and gapless
phases, since we work with finite systems (i.e., with finite bond dimensions),
we do not expect its divergence at criticality. Therefore, we can extract the
value of the correlation length, as proposed in
\cite{chepiga2020floating,chepiga2021lifshitz,chepiga2023eight}, by considering
the correlation function

\begin{equation}    C_{i,j}^{zz}=\langle\sigma^{z}_{i}\sigma^{z}_{j}\rangle-\langle\sigma^{z}_{i}\rangle\langle\sigma^{z}_{j}\rangle,
\end{equation}
where $i$, $j$ denote different sites. This function is related to the
Ornstein-Zenicke form~\cite{ornstein1914influence}

\begin{equation}
	C^{zz}_{i,j}\propto\frac{e^{-|i-j|/\xi}}{\sqrt{|i-j|}}\cos(q|i-j|+\phi_0),
	\label{eq:OZ_equation}
\end{equation}
where the correlation length $\xi$, the wave-vector $q$ and the initial phase
$\phi_0$ are fitting parameters. One example of such fits is shown in Fig.
\ref{fig:xi_fitting}. We discard the oscillations and fit the main slope of the
decay as shown for high precision in the value obtained for $\xi$. Note that the
fit is performed for the logarithm of the correlations, i.e., $\ln
	C^{zz}(x=|i-j|)\approx cte-x/\xi-\ln(x)/2$.

\section{Extraction of the Luttinger Liquid Parameters}\label{ap:
	Luttinger_parameters}

To extract the Luttinger liquid exponent inside the floating phase, we have
fitted the Friedel oscillations of the spin-density profile induced by the open
boundary conditions with Eq.~\ref{eq:Friedel}. Open ends in the chain break
translational invariance and there is a slowly decaying alternating term in the
spin-density
\begin{equation}
	\langle\sigma^{z}_{j}\rangle=\overline{\langle\sigma^{z}_{j}\rangle}+(-1)^{j}\widetilde{\langle\sigma^{z}_{j}\rangle},
\end{equation}
where $\widetilde{\langle\sigma^{z}_{j}\rangle}$ becomes non-zero near the
boundary and decays slowly away from it~\cite{sorensen2007quantum}. In order to
obtain both $\overline{\langle\sigma^{z}_{j}\rangle}$ and
$\widetilde{\langle\sigma^{z}_{j}\rangle}$, we follow the $7$-point formulas
derived in Ref.~\cite{sorensen2007quantum}:
\begin{eqnarray}
	\overline{\langle\sigma^{z}_{j}\rangle}&=&-\frac{15}{496}f(j-3)-\frac{1}{248}f(j-2)+\frac{71}{248}f(j-1)\nonumber\\
	&+&\frac{1}{2}f(j)+\frac{137}{496}f(j+1)+\frac{1}{248}f(j+2)-\frac{1}{31}f(j+3),\nonumber\\
	\widetilde{\langle\sigma^{z}_{j}\rangle}&=&\frac{15}{496}f(j-3)+\frac{1}{248}f(j-2)-\frac{71}{248}f(j-1)\nonumber\\
	&+&\frac{1}{2}f(j)-\frac{137}{496}f(j+1)-\frac{1}{248}f(j+2)+\frac{1}{31}f(j+3).\nonumber\\
\end{eqnarray}

An example of a typical fit is provided in Fig.
\ref{fig:Friedel_oscillation_patterns}. As we see, $20\%$ of sites close to each
end of the chain are discarded and we only fit the middle part to reduce edge
effects.

\bibliography{bibliography}

\end{document}